\documentclass[doublecol]{epl2} 

\usepackage{amsmath}
\usepackage{graphicx}

\title{Determination of the mean center of a region: A physics-based approach}
\shorttitle{Mean center} 

\author{ Dipak Patra \thanks{E-mail: \email{dipak@rri.res.in}}
 }
\shortauthor{Patra \etal}
\institute{                    
  Soft Condensed Matter Group, Raman Research Institute, 
Bangalore 560080, India
}

\abstract{
The mean center of a geographical region, including continents and countries, has been mostly determined to study the trend of population migration, the shift of economic hubs, and the spatial change of extreme climate events. However, the determination of the mean center is a formidable task as it deals with the curvature of the earth's surface.
Here, we report a physics-based model to determine the mean center of a region. Our method provides the analytical expression for the location of the mean center for both flat and curved spaces, such as straight lines, circles, planes, three-dimensional space, cylinders, and spheres. Some of these expressions are often used to compute the center of mass of the physical system. Therefore, the implication of our model in the physical system extends the general validity of the model. Furthermore, we have computed various mean centers of India, such as the geographical, population, and crime centers. We have also assessed the year-wise movement of the crime center and found a spatial trend towards the north.}

\begin{document}
\maketitle

\section{Introduction}
The mean center is ubiquitously associated with the center of mass of a physical system and it is primarily calculated to investigate the dynamics and properties of the system. However, depending on the geographical systems of interest, the mean center is often described as the population center, geographical center, trading center, education center, etc. 
The location of the mean center corresponding to geographical regions, including the globe, continents, countries, and states, is worthwhile to know for conducting various studies concerning economic development \cite{Quah_2011, Kandogan_2014, Grether_2010, Jeleskovic_2023, Sauter_2019, Barreiro_2019}, climate change \cite{Sauter_2019, Barreiro_2019, Yan_2017, Wang_2019, Zeng_2022, Qiao_2023}, population dynamics \cite{Hayford_1902, Barmore_1992, Aboufadel_2006, Plane_2015, Hall_2019, Barreiro_2019} and disease spreading \cite{Deeb_2021} in the regions.  
For the globe, the dynamics of population migration, trading activity, and industrial movement have been mostly assessed by finding the position of the mean center at a time. Similar studies have also been conducted for the continents. Specifically, the mean center is often determined to study the population migration of a country \cite{Barmore_1992, Aboufadel_2006, Plane_2015}.
 It should be noted that the geographical center corresponding to the middle point of a country can be used as the new location of the capital city of the country for security purposes and the convenience of administrative work. Furthermore, knowledge about the location of the center of technological hubs, health infrastructures, and education hubs is crucial to developing the states of a country and eradicating regional disparity.

 However, the determination of the mean center of a region is a non-trivial task as the surface of the earth is not a flat object. 
  Therefore, various types of methods have been developed to determine the mean center of a region \cite{Hayford_1902, Barmore_1992, Aboufadel_2006, Plane_2015}.
 The finding of the mean center of a country dates back to the eighteenth century.
For the first time, the Census Bureau determined the mean center based on the survey data corresponding to the population of the USA. The position of the population center was determined as
  \begin{eqnarray*}
 \bar{\phi}&=&\frac{\sum_i w_i \phi_i}{\sum_j w_i}~,\\
 \bar{\lambda}&=&\frac{\sum_i w_i \lambda_i \cos\phi_i}{\sum_j w_j \cos\phi_j}~,
 \end{eqnarray*}
 where the latitude and longitude are denoted by $\phi$ and $\lambda$, respectively, and the population at a point is represented by the positive variable $w$.
 Barmore pointed out that the Census Bureau employed  Sanson-Flamsteed or sinusoidal projection of Earth on a flat two-dimensional space \cite{Barmore_1992}.
  He argued various flaws of this method and discussed the properties of the mean center. Barmore also calculated the mean center for the population by using Cartesian coordinates of each point $(x,y,z)$ as follows
 \begin{eqnarray*}
 x&=&\frac{\sum_i w_i x_i}{\sum_j w_i}~,\\
 y&=&\frac{\sum_i w_i y_i}{\sum_j w_j}~,\\
 z&=&\frac{\sum_i w_i z_i}{\sum_j w_j}~.
 \end{eqnarray*}
The latitude and longitude of the center of the population can be found as
\begin{eqnarray*}
\phi &=& \sin^{-1}\frac{z}{R}~,\\
\lambda &=& \tan^{-1} \frac{y}{x}~,
\end{eqnarray*}
where the radius of the earth is denoted by $R$.
 It should be noted that the exact three-dimensional location of this center is always beneath the surface of the earth as it follows $x^2 +y^2 +z^2 <R^2$.
Therefore, this center is not accessible from the surface of the earth. Because of this limitation, Barmore discarded this method and employed an iterative method for the determination of the center using an azimuthal equidistant map \cite{Barmore_1992}.

 Aboufadel \textit{et al}. argued that the using of different methods projecting the surface of the earth on a two-dimensional plane leads to different locations of the mean center for the same data, and hence, the center depends on the choice of projection methods \cite{Aboufadel_2006}. They also pointed out the limitation of the iterative method proposed by Barmore. Therefore, they proposed a force-balancing method to calculate the center of the population. 
By balancing the forces acting towards the earth center corresponding to each point, they computed a unit vector whose components are given by
  \begin{eqnarray*}
 \bar{x}&=&\frac{\sum_i w_i x_i}{|\sum_i w_j \vec{r}_j|}~,\\
 \bar{y}&=&\frac{\sum_i w_i y_i}{|\sum_i w_j \vec{r}_j|}~,\\
 \bar{z}&=&\frac{\sum_i w_i z_i}{|\sum_i w_j \vec{r}_j|}~,
 \end{eqnarray*}
 where $\vec{r}_i=(x_i,y_i,z_i)$ denotes the position of the $i$-th point on the surface of earth.
 From the unit vector, they calculated the location of the center of the population as
 \begin{eqnarray}
\phi &=& \sin^{-1}\bar{z}~,\\
\lambda &=& \tan^{-1} \frac{\bar{y}}{\bar{x}}~.
\end{eqnarray}
 However, this approach exactly gives the same location obtained from the Cartesian method employed by Barmore. 
 
Nevertheless, Barmore also speculated that one can find the mean center by minimizing the sum of the square geodesic distances of the points on the sphere \cite{Barmore_1992}. 
 Recently, this idea has been employed to determine the mean center \cite{Plane_2015}.
 Not only is this procedure complex, but it also requires large numerical resources. According to Barmore, this method becomes even more complex for the ellipsoidal surface.
 
Despite these methods, another method has been utilized to study the spatial change of climate events and cropland \cite{Yan_2017, Wang_2019, Zeng_2022, Qiao_2023}, where the mean center is defined as 
   \begin{eqnarray*}
 \bar{\phi}&=&\frac{\sum_i w_i \phi_i}{\sum_j w_i}~,\\
 \bar{\lambda}&=&\frac{\sum_i w_i \lambda_i }{\sum_j w_j }~.
 \end{eqnarray*}
 In these studies, the positive variable $w$ is associated with the quantities either the density of pollutants or the amount of cropland.
However, this definition does not deal with the curvature of the earth and is also limited according to the properties of the mean center, as argued by Barmore \cite{Barmore_1992}.
 
 \begin{figure}[ht]
   \includegraphics[clip=true,width=\columnwidth]{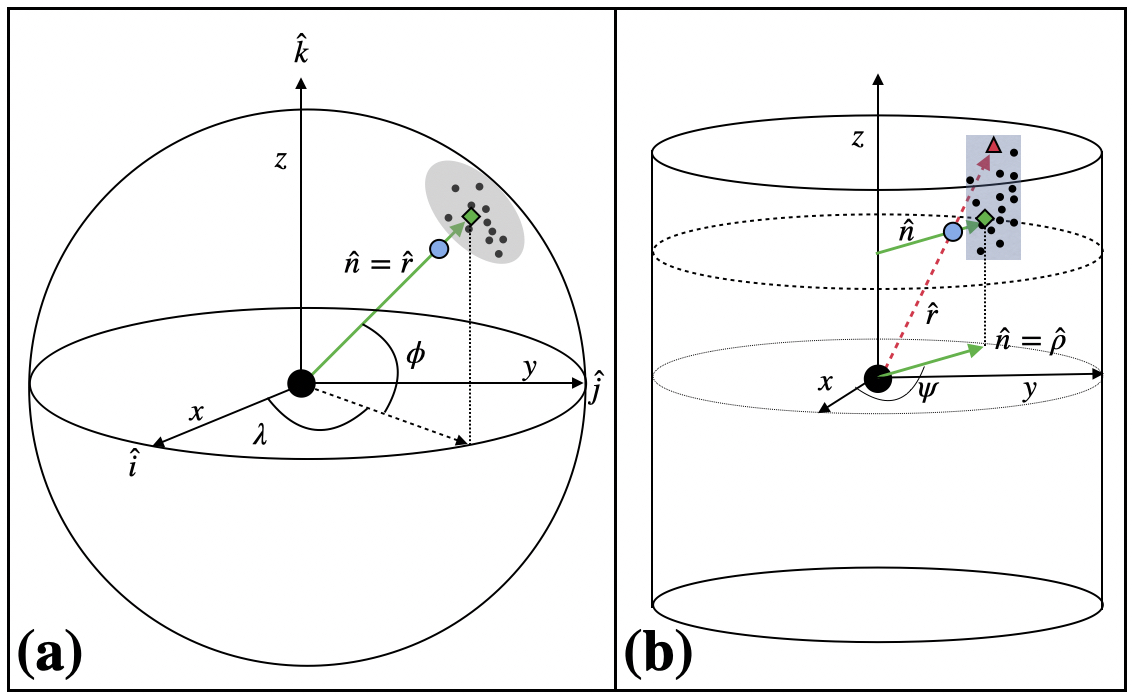}
   \caption{
   Schematic representation of the mean center for (a) sphere and (b) cylinder denoted by the blue colored circle using the Cartesian method. The radial direction and normal unit vector to the surface are denoted by $\hat{r}$ and $\hat{n}$, respectively. For the spherical surface, $\phi$ and $\lambda$ are the latitude and longitude of a point, respectively, whereas $\psi$ is the azimuthal angle of a point on the cylindrical surface.
The actual location of the mean center on the surface is marked by the green-colored diamond symbol. The projection along the radial direction provides a different point denoted by the red-colored triangle for the cylindrical surface.  }
   \label{fig:MCSchematic}
\end{figure}

 Among these procedures described above, the force balancing method for the determination of the mean center proposed by Aboufadel \textit{et al.} has been used in most of the studies
 because of the apparent simplicity of the calculation \cite{Hall_2019, Quah_2011, Kandogan_2014, Deeb_2021}.
 However, this force balancing method has also some limitations that are discussed as follows.
  In principle, one can think that at first, they calculated the mean center using Cartesian coordinates. The center is expected to be located inside of the earth. To resolve the issue about the inaccessibility of the center point raised by Barmore, the point is then projected on the surface of the earth along the radial direction as shown in figure~\ref{fig:MCSchematic}(a). Therefore, the projected location is taken as the true location of the mean center. However, the projection along the radial direction leads to a different location of the mean center for the cylindrical surface, as shown in figure~\ref{fig:MCSchematic}(b). 
Therefore, the general validation of the projection is still lacking.
  The force balancing method is found to be easy for the spherical surface. However, this method is conceptually difficult even for the cylindrical surface. Therefore, the implication of this method to find the mean center for an arbitrary surface such as an ellipsoid is a formidable task. It should be noted that the Earth is not a perfectly spherical object and hence this method is not warranted for the closely ellipsoidal earth.

 Here, we propose a new method to identify the mean center for an arbitrary parametrizable surface using a particle-based physics model. The mean centers for one, two, and three-dimensional flat spaces are calculated. We also determine the mean center for cylindrical and spherical surfaces. Furthermore, we compute the geographical, population, and crime centers for India as a test case for the proposed method. 

 \section{Method}
 In this section, we describe our physics-based method to determine the mean center of the provided locations of a region. For the geographical systems, the locations can be associated with the positions of various entities such as villages, towns, cities, states, and countries.
We assume that the locations of the entities are fixed on the space, and therefore, the entities are treated as frozen point particles. 
 It is also assumed that the point particles are connected to the mean center with virtual springs. Therefore, the elastic energy between $i$-th particle and the mean center can be written as $u_i=\frac{1}{2} w_i |\vec{r}-\vec{r}_i|^2 $
where  $\vec{r}=(x,y,z)$ and $\vec{r}_i=(x,y,z)$ denote the position of the mean center and $i$-th particle in the Cartesian coordinate frame, respectively and the positive coefficient $w_i$ is the spring constant. The spring constant $w$ can be associated with various quantities depending on the system of interests, such as the mass of particles for the physical system, the number of active patients for disease spreading, the density of pollutants for climate analysis, etc. In addition, to find the population center of a region, the weighted population number is considered to be equal to $w$. 
It should be noted that for the region containing only a single particle, the mean center is supposed to be at the location of the single particle. This is ensured by the energy being equal to zero for the same location of the particle and the mean center. 
 The total elastic potential energy of the region containing $N$ particles is expressed as
 \begin{equation*}
 U_{el}=\frac{1}{2}\sum_{i=1}^N w_i |\vec{r}-\vec{r}_i|^2 ~.
 \end{equation*}
We also assume that the region is a curved surface embedded in three dimensions, and the particles of the region are on the surface. Therefore, the mean center is expected to lie on the surface, giving rise to a constraint 
 \begin{equation}\label{eqn:surface}
 g(\vec{r})=f(x,y,z) ~.
 \end{equation}
 However, this constraint can be treated as an additional potential energy term 
 $U_c= \mu g(\vec{r})$,
 where $\mu$ is the Lagrange undetermined multiplier. 
  Therefore, the total potential energy of the system is given by
 \begin{equation*}
U= U_{el}+U_c=\frac{1}{2}\sum_{i=1}^N w_i |\vec{r}-\vec{r}_i|^2 + \mu g ~.
 \end{equation*}
 The total potential energy deals with the distances between the center and frozen particles in the Cartesian coordinate system. It should be noted that the Cartesian distances among the particles of physical systems have been computed in the molecular dynamics simulations investigating the collective movement of the particles on curved surfaces such as cylinders \cite{Hindes_2020}, corrugated tubes \cite{Ai_2019}, ellipsoids \cite{Ehrig_2017, Alaimo_2017}, and spheres \cite{Sknepnek_2015, Alaimo_2017, Henkes_2018, Hindes_2020}. 
Following the literature, we compute the Cartesian distances between the center and frozen particles of the system irrespective of the flat or curved spaces.
To represent this system as a physical system, we consider a test particle of mass $m$ at the position of the mean center $\vec{r}=(x,y,z)$. This test particle moves with a velocity $\vec{v}=\dot{\vec{r}}$ on the surface described by equation~(\ref{eqn:surface}). Therefore, the kinetic energy of the test particle is given by 
\begin{equation*}
 T=\frac{1}{2} m |\dot{\vec{r}}|^2~,
 \end{equation*}
and the Lagrangian of the system is defined as $L=T-U$.
 The equation of motion of the test particle is given by the Euler-Lagrange equation
 \begin{equation*}
\frac{d}{dt}\frac{\partial L}{\partial \dot{q}_\alpha}  - \frac{\partial L}{\partial q_\alpha} =0~,
 \end{equation*}
 where $q_\alpha =\{x,y,z \} $ is the generalised coordinate \cite{Goldstein_2001}. 
 From the Euler-Lagrange equation, one can obtain the component-wise equations
 \begin{eqnarray*}
 m\ddot{x}- \sum_{i=1}^N w_i (x-x_i) -\mu \eta_x &=& 0~,\\
  m\ddot{y}- \sum_{i=1}^N w_i (y-y_i) -\mu \eta_y &=& 0~,\\
   m\ddot{z}- \sum_{i=1}^N w_i (z-z_i) -\mu \eta_z &=& 0 ~,
 \end{eqnarray*}
where $\vec{\eta}= (\eta_x,\eta_y,\eta_z)=\nabla g=(\frac{\partial g}{\partial x}, \frac{\partial g}{\partial y}, \frac{\partial g}{\partial z} )$ is a normal vector to the surface.
 All of these equations can be rewritten in a single equation as
 \begin{equation} \label{eqn:g_newton}
m\ddot{\vec{r}}= \vec{F} + \mu \vec{\eta} ~,
 \end{equation} 
 where 
 \begin{equation} \label{eqn:force}
\vec{F} = \sum_{i=1}^N w_i (\vec{r}-\vec{r}_i) 
 \end{equation}
 represents the total force acting upon the test particle due to the elastic springs.
 The constant $\mu$ can be determined as follows.
 By taking the dot product of $\eta$ and the vector of both sides of the equation~(\ref{eqn:g_newton}), we obtain 
  \begin{equation} \label{eqn:lambda}
 \mu = \frac{ m\ddot{\vec{r}}\cdot \eta - \vec{F}\cdot \eta }{ |\vec{\eta}|^2 } ~.
 \end{equation} 
  From the constraint equation~(\ref{eqn:surface}), one can obtain the auxiliary equation as
   \begin{equation*}
\frac{d g(\vec{r})}{dt}=\vec{\eta}\cdot\dot{\vec{r}}=\vec{\eta}\cdot\vec{v}=0~.
 \end{equation*}
 The second derivative of $g(\vec{r})$ with respect to time provides the additional constraint equation as
  \begin{equation*}
\frac{d^2 g}{dt^2}  = \vec{\eta}\cdot\dot{\vec{v}}+\dot{\vec{\eta}}\cdot\vec{v} =0 ~.
 \end{equation*}
 From this equation, we obtain
$\vec{\eta}\cdot\dot{\vec{v}} =-\dot{\vec{\eta}}\cdot\vec{v}$.
By plugging the expression $\vec{\eta}\cdot\dot{\vec{v}} = \vec{\eta}\cdot\ddot{\vec{r}}$ into equation~(\ref{eqn:lambda}), one can find the Lagrange multiplier constant
 \begin{equation*}
\mu = -\frac{m\dot{\vec{\eta}}\cdot\vec{v} + \vec{\eta} \cdot \vec{F} }{|\vec{\eta}|^2}.
 \end{equation*}
 Therefore, the general equation of motion of the test particle is rewritten as
 \begin{equation}\label{eqn:newton}
m\ddot{\vec{r}}= m\dot{\vec{v}}= \vec{F} -\vec{\eta} \frac{m\dot{\vec{\eta}}\cdot\vec{v} + \vec{\eta} \cdot \vec{F} }{|\vec{\eta}|^2} ~.
 \end{equation}
It should be noted that the location of the mean center of the region is determined by the equilibrium position of the test particles. At the equilibrium states, the acceleration ($\dot{\vec{v}}$) and the velocity ($\vec{v}$) of the test particle is expected to be zero. Therefore, by putting $\dot{\vec{v}}=0$ and $\vec{v}=0 $ in equation~(\ref{eqn:newton}), one can obtain the condition for the mean center as
 \begin{equation*}
  \vec{F} -\vec{\eta} \frac{ \vec{\eta} \cdot \vec{F} }{|\vec{\eta}|^2} = 0 ~.
 \end{equation*}
 The equilibrium condition can be rewritten as 
  \begin{equation}\label{eqn:FinalCM}
  \vec{F} -(\vec{F}\cdot\hat{n})\hat{n} = 0 ~,
  \end{equation}
 where $\hat{n}=\frac{\vec{\eta}}{|\vec{\eta}|}$ is the unit normal vector to the surface at the position (i.e. a point $\vec{r}$ on the surface) of the test particle.

 \section{ Mean Centers for various space }
 Here, we find the expression for the mean center for various spaces, including flat and curved spaces. 
For a straight line, the vectors $\vec{r} - \vec{r}_i$  are always on the line, and hence, they are perpendicular to $\hat{n}$ (i.e. normal unit vector to the line), giving rise to $\vec{F}\cdot\hat{n}=0$. The equilibrium condition in equation~(\ref{eqn:FinalCM}) therefore becomes
 $\vec{F}=0$ ,
 and then the mean center is defined as 
 \begin{equation}
 \vec{r}=\frac{\sum_i w_i \vec{r}_i }{\sum_i w_i}~.
 \end{equation}
 
For a plane, the vectors
$\vec{r} - \vec{r}_i$ are always in the plane, and they are perpendicular to the normal unit vector ($\hat{n}$) of the plane. Following a similar procedure to the straight rod, one can find the position of the mean center
\begin{equation}
 \vec{r}=\frac{\sum_i w_i \vec{r}_i }{\sum_i w_i} ~.
 \end{equation}
 
 Similarly for the three-dimensional flat space, one can obtain the mean center as
\begin{equation}
 \vec{r}=\frac{\sum_i w_i \vec{r}_i }{\sum_i w_i} ~.
 \end{equation}
 
For a circular line, the line is parametrized as
$g(x,y)=x^2+y^2 -R^2$, where the radius of the circular arc is denoted by $R$. The unit normal vector is given by
$\hat{n}=(\frac{x}{\sqrt{x^2 +y^2}}, \frac{y}{\sqrt{x^2 +y^2}}) =\hat{r}$. The total force can contain the tangential and normal forces, and it can be represented  as $
\vec{F}= (\vec{F}\cdot \hat{r}) \hat{r} + (\vec{F}\cdot \hat{\psi})  \hat{\psi}$, where the tangential direction is denoted by a unit vector $\hat{\psi}$. However, the tangential force becomes zero due to the equilibrium condition described in equation~(\ref{eqn:FinalCM}).
Furthermore, from this equilibrium condition, one can find 
\begin{equation}
\vec{P}=(\vec{P}\cdot \hat{r}) \hat{r}~,
 \end{equation}
 where $\vec{P}=\sum_i w_i \vec{r}_i$ denotes the first moment. Therefore, the tangential component of $\vec{P}$ equals zero. The location of the mean center is given by
\begin{equation}
 \hat{r}=\frac{\vec{P}}{|\vec{P}|}=\frac{\sum_i w_i \vec{r}_i }{|\sum_i w_i \vec{r}_i|} ~.
 \end{equation}
 
For a cylindrical surface, the surface is described as $g(x,y)=x^2+y^2 -R^2$ and $z=z$, where the radius of the cylinder is denoted by $R$. The position vector $\vec{r}(x,y,z)$ of a particle is represented in the cylindrical polar coordinate ($\rho, \psi, z$) as
 $\vec{r}=\rho\hat{\rho} +z \hat{k}$.
  The normal unit vector to the cylindrical surface is found as $\hat{n}=(\frac{x}{\sqrt{x^2 +y^2}}, \frac{y}{\sqrt{x^2 +y^2}},0) =\hat{\rho}$. The total force can be decomposed as 
$\vec{F}= \sum_{i=1}^N w_i (\vec{\rho} -\vec{\rho_i}) + \sum_{i=1}^N w_i (z -z_i)\hat{k}$. 
  The equilibrium condition in equation~(\ref{eqn:FinalCM}) gives $ \sum_{i=1}^N w_i (z -z_i) = 0 $ as $\hat{\rho}\cdot \hat{k}=0$ . Thus, one obtains the $z$ component of the position vector of the mean center as 
\begin{equation}
z=\frac{\sum_i w_i z_i}{\sum_j w_j} ~.
 \end{equation}
Furthermore, the equilibrium condition in equation~(\ref{eqn:FinalCM}) provides
\begin{equation*}
\vec{P}=(\vec{P}\cdot \hat{\rho}) \hat{\rho}~,
 \end{equation*}
 where $\vec{P}=\sum_i w_i \vec{\rho}_i$. 
 Therefore, $\hat{\rho}$ corresponding to the mean center is given by
\begin{equation}
 \hat{\rho}=\frac{\vec{P}}{|\vec{P}|}=\frac{\sum_i w_i \vec{\rho}_i }{|\sum_i w_i \vec{\rho}_i|} ~.
 \end{equation}
 From $\hat{\rho}$ one can find the $x$ and $y$ positions of the mean center corresponding to the cylindrical surface.
 
 For the spherical surface, the surface is described as $x^2 +y^2 +z^2 = R^2$, where $R$ is the radius of the sphere. The normal of the sphere is along the radial direction, i.e. $\hat{n}=\hat{r}$. The equilibrium condition in equation~(\ref{eqn:FinalCM}) gives
\begin{equation}
 \hat{r}=\frac{\vec{P}}{|\vec{P}|}=\frac{\sum_i w_i \vec{r}_i }{|\sum_i w_i \vec{r}_i|}~,
 \end{equation}
 where $\vec{P} = \sum_i w_i \vec{r}_i$ is the first moment.
 From the unit vector $\hat{r} =(\bar{x},\bar{y},\bar{z})$, the latitude and longitude of the mean center can  be described as
 \begin{eqnarray}
\phi &=& \sin^{-1}\bar{z}~,\\
\lambda &=& \tan^{-1} \frac{\bar{y}}{\bar{x}}~.
\end{eqnarray}
For other space
 such as ellipse and ellipsoid, the normal unit vector $\hat{n}$ needs to be found first. In these cases, $\hat{n}$ is not along the radial direction $\hat{r}$ (i.e. $\hat{n} \neq \hat{r}$). Finding the analytical solution of the equation~(\ref{eqn:FinalCM}) is a formidable task.  Therefore, the equation~(\ref{eqn:FinalCM}) has to be solved numerically for the determination of the position of the mean center.

\section{Discussion}
Our proposed method provides the exact analytical expression of the mean centers corresponding to the straight lines, planes, and three-dimensional space which are often utilized to determine the center of mass of physical systems. For the first time, we find the expression of the mean center for the circular arc and the cylindrical surface, which can be used for the physical systems.  For the spherical surface, our method gives the same result obtained from the force balancing method \cite{Aboufadel_2006}. This agreement extends the general validity of our model. In principle, the mean center for the geographical system is determined by assuming the spherical surface of the earth. It is a difficult task to incorporate the exact shape of the earth in the calculation of the mean center. In this case, our method easily takes account of the ellipsoidal surface of the earth, but the mean center has to be determined numerically. Therefore, our method is a robust one that can be utilized for flat spaces as well as curved surfaces.
It should be noted that 
our method is not only applicable to geographical systems but also to physical systems. For example, active particles moving in flat spaces or on curved surfaces, such as spheres and ellipsoids, are known to exhibit flocking behavior like fish school, bird swarm, and locus swarm \cite{Sknepnek_2015, Ehrig_2017, Ai_2019}.
The center of mass of the flocking state needs to be assessed to understand the dynamics of this state and it can be determined easily by our method.
Therefore, the implication in geographical and physical systems makes our method a general kind.

Most of the methods involving projection maps or 
geodesic distances raise a fundamental issue when the points of equal strength ($w$) are evenly distributed over the spherical surface. In this condition, one can not determine the location of the mean center on the surface of the sphere. However, these methods give the specific set of values of $\bar{\phi},\bar{ \lambda}$, which leads to the artifact of the system (see the expression in the Introduction section for the mean center used by the Census Bureau).
 On the other hand, our method using Cartesian coordinate gives $\vec{P}=0$, which implies that the mean center can not be projected on the surface and hence it can not be found. If one insists on finding the mean center in this case, then the mean center would be the center of the sphere.
\begin{largetable}
\caption{Different types of the mean center for India. All the centers are in any of the two states, namely Madhya Pradesh (MP) and Maharashtra (MH). }\label{tab:MC}
\begin{tabular}{p{2 in}  p{1in} p{1in} p{1.5in} }
Mean center & Latitude & Longitude & Place \\
\hline
Geographical center & $22.5182^\circ$ N & $79.1734^\circ$ E & Dabri (MP)\\
Population center   & $22.1141^\circ$ N & $79.0038^\circ$ E & Chhindwara (MP) \\
Crime center (2020) & $20.6402^\circ$ N & $78.1514^\circ$ E & Babhulgaon (MH) \\
Crime center (2021) & $21.8027^\circ$ N & $78.0886^\circ$ E & Junapani (MP) \\
Crime center (2022) & $21.7146^\circ$ N & $78.1958^\circ$ E & Chandora Dam (MP) \\
\hline
\end{tabular}
 \end{largetable}
  \begin{figure*}[ht]
  \centering
 \includegraphics[height=7.cm, width=16cm,keepaspectratio]{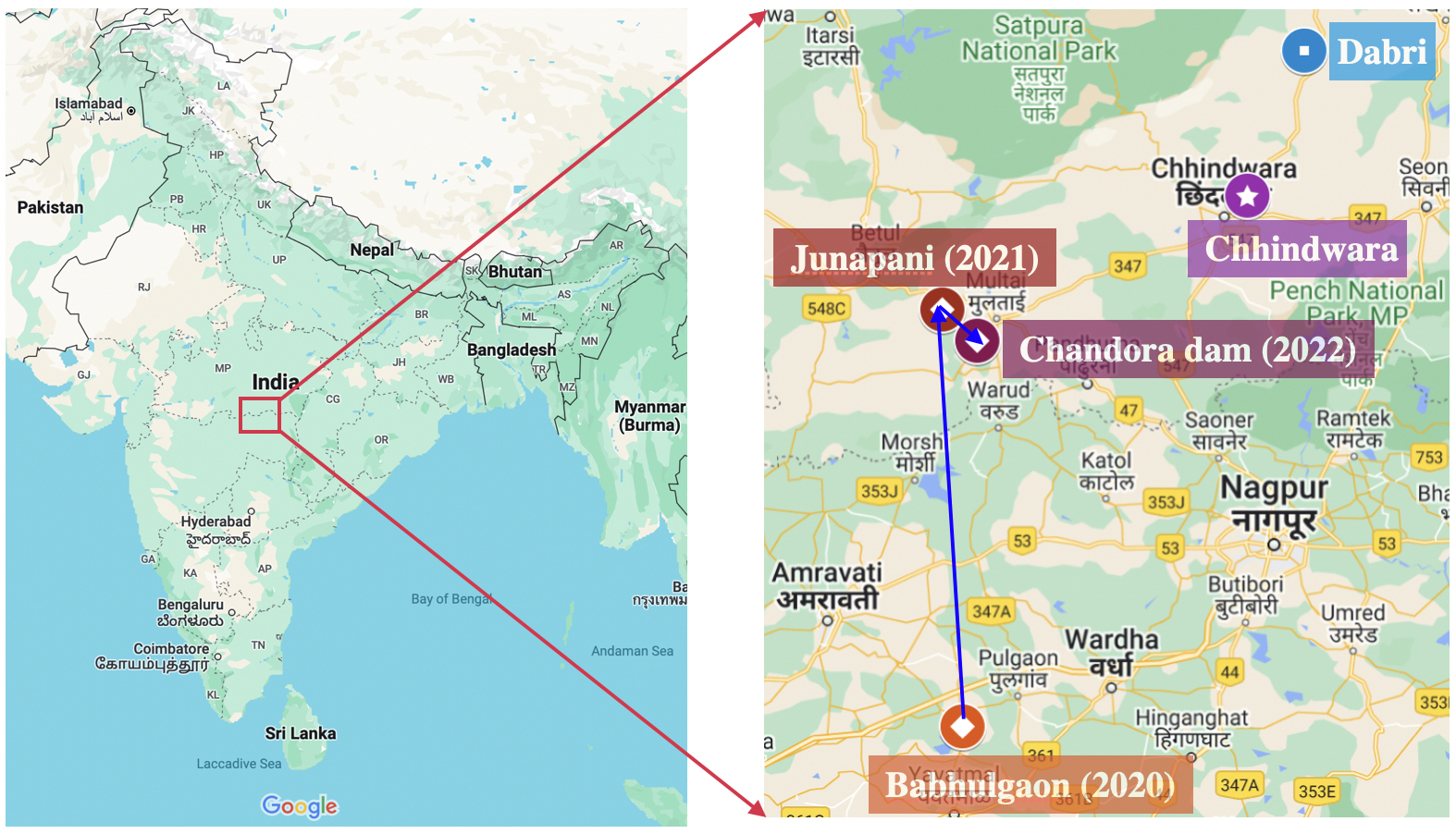}
   \caption{ Displaying the map of India. The right pan is the magnified view of the marked region enclosed by a red box. The location of various mean centers including geographical, population, and crime centers of India are indicated in the right pan. The year-wise movement of the crime center is depicted by the blue arrows. }
    \label{fig:MC_India}
\end{figure*}
 \section{Test Case}
We have computed the mean centers for India using the proposed physics-based model. For these computations, we have assumed the Earth is a perfect spherical object.
Furthermore, all the states and union territories are considered and also the locations of the capital places are chosen as the locations of the states respectively. The latitude and longitude of these capitals are obtained from Google. Population data for the states are collected from the Unique Identification Authority of India report \cite{UIDAI_2020}. The Indian Penal Code (IPC) crime data are obtained from the National Crime Records Bureau report entitled ``Crime In India" for the subsequent years 2020, 2021 and 2022 \cite{NCRB_2022}. 
In table~\ref{tab:MC}, we present the location of these mean centers for the country India. For the computation of the geographical center, equal importance is given to all the states and union territories, which is ensured by the same value of weight $w$ for these regions. Therefore, the geographical center can be taken as the middle point of the country. It is found that the geographical center is located near Dabri in Madhya Pradesh state (see figure~\ref{fig:MC_India} ). For the determination of the population center, the weight $w$ is assumed to be equal to the population number corresponding to the zones. The population center is also found to be located in Madhya Pradesh state, which is also near the geographical center.
Similarly, we have computed the crime center where the weight $w$ is assumed to be equal to the number of IPC crimes corresponding to each region. The crime center is found to be located near Babhulgaon in Maharastra state for the year 2020. It should be noted that the crime center shows a large movement towards the north for the year 2021 and is located near Junapani in  Madhya Pradesh state as shown in the right pane of figure~\ref{fig:MC_India}. 
For 2022, the center moves slightly towards the south and is found near the place Chandora Dam in Madhya Pradesh. 
Therefore, the year-wise location of the crime center provides the spatial trend of the crime pattern across the country. However, the study over the three years is not sufficient for understanding the dynamics of the crime center. Therefore, further study over a long period is needed to find the correct trend of the crime center.
 \section{Summary}
 In summary, we have developed a physics-based model to determine the mean center of a region. Employing this model, we have found the analytical expression for the location of the mean center for both flat and curved spaces such as straight lines, circles, planes, three-dimensional space, cylinders, and spheres. Some of the expressions are commonly used to determine the center of mass of the physical system. Using the expression for the sphere, we have computed various mean centers of the country India, such as geographical, population, and crime centers. Furthermore, we have assessed the trend of the year-wise movement of the crime center.

\acknowledgments
We thank Subham Ghosh, Saikat Shyamal, Arpita Das, and Anindya Chowdhury for reading and providing critical comments to improve the manuscript.

\textit{Data availability statement}: 
Data are available on request to the author.

\bibliographystyle{eplbib}

\end{document}